\documentclass[final,3p,times]{elsarticle}

\usepackage{amsmath}
\usepackage{amssymb}
\usepackage{geometry}
\AtBeginDocument{\fontsize{11pt}{14.5pt}\selectfont}
\usepackage{titlesec}

\titleformat{\section}
  {\normalfont\bfseries\fontsize{12pt}{14pt}\selectfont}
  {\thesection}
  {0.8em}
  {}

\titleformat{\subsection}
  {\normalfont\bfseries\fontsize{11pt}{13pt}\selectfont}
  {\thesubsection}
  {0.8em}
  {}

\titleformat{\subsubsection}
  {\normalfont\bfseries\fontsize{11pt}{13pt}\selectfont}
  {\thesubsubsection}
  {0.8em}
  {}

\usepackage[section]{placeins}

\usepackage[
    colorlinks=true,
    linkcolor=blue,
    urlcolor=blue,
    citecolor=blue
]{hyperref}

\begin{document}

\begin{frontmatter}

\title{PhonoMC: Occupation-based deviational Monte Carlo for phonon transport with temperature-dependent scattering}

\author[1]{Shixian Liu\corref{cor1}}
\ead{lyu@bmstu.ru}

\author[1]{Fei Yin}

\author[2]{Gang Wang}

\author[3]{Bin Liu}

\author[1]{Ge Zhang}

\author[1]{Alexander A. Barinov\corref{cor1}}
\ead{barinov@bmstu.ru}

\author[2]{Ke Xu\corref{cor1}}
\ead{kickhsu@gmail.com}

\cortext[cor1]{Corresponding authors}

\address[1]{
Department of Thermophysics,
Bauman Moscow State Technical University,
Moscow 105005, Russia
}

\address[2]{
College of Physical Science and Technology,
Bohai University,
Jinzhou, China
}

\address[3]{
Key Laboratory for Thermal Science and Power Engineering of Ministry of Education,
Department of Engineering Mechanics,
Tsinghua University,
Beijing 100084, China
}

\begin{abstract}
Nanoscale self-heating involves phonon transport across confined geometries, material interfaces, and temperature fields over which the scattering rates can vary substantially. We develop PhonoMC, an occupation-based deviational Monte Carlo method for solving the phonon Boltzmann transport equation within the relaxation-time approximation. A fixed equilibrium state is retained as the deviational reference, while the local temperature reconstructed from the represented energy is used to evaluate mode-dependent scattering rates. Collisions are updated with a separately determined relaxation temperature to conserve energy over each time step, and prescribed lattice heating is introduced by changing carrier occupations rather than continuously adding computational particles. For cross-plane transport through a 100-nm Si film, the deviational formulation reproduces the full-population heat flux while reducing its standard deviation by a factor of approximately 3.3 at \(\Delta T=100\)~K with \(10^5\) carriers. In contrast, keeping the scattering rates fixed at 300~K overestimates the heat flux by 23.7\% at \(\Delta T=250\)~K. Calculations of Si thin films distinguish finite-length effects from surface-scattering suppression, and Si/3C-SiC bilayers are used to examine interfacial thermal resistance. The method is further applied to localized heating in FinFET-like structures, where replacing the lower Si substrate with higher-conductivity 3C-SiC leads to a higher hotspot temperature because of the additional resistance associated with the confined Si region and the Si/SiC interface. These results show that a fixed deviational reference can be combined with local temperature-dependent scattering and sustained heat deposition in a mode-resolved Monte Carlo description of nanoscale thermal transport.
\end{abstract}


\end{frontmatter}

\section{Introduction}

Heat dissipation has become an increasingly important constraint in semiconductor electronics as device dimensions shrink and heat generation becomes more spatially localized~\cite{pop_heat_2006,pop_monte_2005}. Localized Joule heating can generate nanoscale hotspots whose temperatures are determined not only by the amount of heat produced but also by the available pathways for heat removal~\cite{sinha_non-equilibriumphonon_2006}. When characteristic device dimensions, film thicknesses, or hotspot sizes become comparable to phonon mean free paths, heat transport becomes quasiballistic and can no longer be described reliably by a local Fourier law using bulk thermal conductivity~\cite{chen2021non-fourier,chen_ballistic-diffusive_2001,minnich2011thermal,kaiser_thermal_2017,Anufriev2018a,razavi_review_2016}. This issue is particularly relevant to fin field-effect transistors (FinFETs), where heat is transported through strongly confined three-dimensional structures~\cite{sverdrup_measurement_2001,xu2023quantitative}. Replacing part of the substrate with a material of higher bulk thermal conductivity may improve heat spreading, but the resulting heterointerface introduces an additional resistance that can alter the overall heat-removal pathway.

The phonon Boltzmann transport equation (BTE) provides a kinetic description of heat transport beyond the Fourier regime. In the diffusive limit, frequent scattering drives the phonon population toward local equilibrium and recovers Fourier behavior at the macroscopic scale. Under nanoscale confinement, however, phonons with different frequencies, group velocities, and relaxation times can depart from equilibrium by very different amounts~\cite{chen_ballistic-diffusive_2001,mazumder2001monte,kaiser_thermal_2017,mcgaughey_size-dependent_2011}. Resolving these differences requires spectral and directional information at the mode level~\cite{jia_effective_2025,liu2024determination,anufriev_impact_2023,liu_quantifying_2025,wu_isotope_2024}; gray or strongly averaged descriptions may otherwise introduce errors that depend on geometry and transport regime~\cite{li_ballistic-diffusive_2020,shen_near-junction_2023}. The numerical challenge is therefore to retain mode-resolved transport physics while treating nonuniform temperature fields in geometries large and complex enough to represent nanoscale devices.

Several numerical strategies have been developed for this purpose, including discrete ordinates methods~\cite{mittal_hybrid_2011,mittal_generalized_2011,loy_coupled_2015,hu_giftbte_2023,jia_effective_2025,sheng_multiscale_2025}, lattice Boltzmann formulations~\cite{nabovati_lattice_2011,guo_lattice_2016}, gas-kinetic and synthetic iterative schemes~\cite{liu_unified_2025,zhang_synthetic_2025,zhang_discrete_2025}, and Monte Carlo (MC) approaches~\cite{peterson1994direct,mazumder2001monte,lacroix_monte_2005,peraud_efficient_2011}. Deterministic solvers avoid stochastic noise through direct phase-space discretization, and recent acceleration strategies have substantially extended their range of applicability. Mode-resolved packages such as almaBTE~\cite{carrete_almabte_2017}, GiftBTE and its subsequent developments~\cite{hu_giftbte_2023,jia_effective_2025,sheng_multiscale_2025}, and JAX-BTE~\cite{shang_jax-bte_2025} have further improved access to BTE calculations, while nonlinear synthetic iterative methods have been developed specifically for hotspot transport with temperature-dependent properties~\cite{zhang_synthetic_2025}. MC methods offer a complementary description based on sampled carrier trajectories and are particularly convenient when transport involves complex boundaries, interfaces, or spatially localized heat sources.

Phonon MC methods have evolved from models based on simplified dispersions~\cite{peterson1994direct} to frequency- and polarization-resolved formulations~\cite{mazumder2001monte,lacroix_monte_2005,hao_frequency-dependent_2009,kukita_monte_2013} and, more recently, full-band approaches that retain wave-vector- and branch-resolved phonon properties~\cite{peng_monte-carlo_2022,silva_monte_2024,shen_near-junction_2023}. Deviational and variance-reduced formulations improve statistical efficiency by treating a chosen equilibrium state analytically and sampling only departures from that state~\cite{peraud_efficient_2011,peraud2012alternative,peraud_adjoint-based_2015,forghani_reconstruction_2016,pathak_mcbte_2021,zhou_effect_2026}. Related ray-tracing tools, including P-TRANS and FreePATHS, have been developed to analyze phonon trajectories and surface scattering in geometrically complex nanostructures~\cite{shao_p-trans_2022,anufriev_ray_2020}. These developments have enabled MC studies of films and nanowires~\cite{lacroix_monte_2005,lacroix_monte_2006}, porous and patterned structures~\cite{jeng2008Model,hao_frequency-dependent_2009,wolf_thermal_2014,shao_p-trans_2022,anufriev_ray_2020}, and transistor-like systems containing localized heat sources~\cite{shen_near-junction_2023,chen_coupled_2023,li_transient_2024}.

For transport near a prescribed reference temperature, evaluating phonon scattering rates at that temperature is consistent with a linear-response treatment. This approximation becomes restrictive when localized heating produces a substantial temperature rise, because the evolving temperature field also modifies the mode-dependent scattering rates and hence the subsequent transport dynamics~\cite{xie_temperature-corrected_2025}. This distinction is important in deviational MC. The equilibrium reference introduced for variance reduction does not require the material properties to remain fixed at the reference temperature. Likewise, subtracting an equilibrium reference is not equivalent to linearizing the BTE; linearization requires an additional approximation to the equilibrium distribution or collision operator~\cite{peraud_efficient_2011,peraud2012alternative}. A finite-temperature deviational formulation must therefore reconstruct the local thermal state independently of the reference distribution and use that state consistently when updating temperature-dependent scattering.

Mode-dependent relaxation introduces a second numerical issue. The temperature reconstructed from the total local energy defines the thermodynamic state of a cell and can therefore be used to evaluate temperature-dependent lifetimes. It does not, however, generally provide the equilibrium target required for an energy-conserving mode-dependent RTA collision, because different phonon modes relax at different rates. A separate rate-weighted relaxation temperature must instead be obtained from the collision-energy constraint. The reference temperature, the local temperature used to evaluate scattering properties, and the conserving relaxation temperature thus play different numerical roles. Localized heating creates an additional practical difficulty: direct source injection can continuously increase the number of computational carriers during sustained heating. These two issues motivate the occupation-based formulation developed here.

We introduce PhonoMC, a full-band occupation-based deviational MC framework for finite-temperature phonon transport with localized heating. Computational carriers retain fixed phase-space weights while their modal occupations evolve relative to a fixed equilibrium reference. The local temperature is reconstructed from the represented energy and used to update the mode-dependent scattering rates. During the collision step, a separate relaxation temperature is obtained from discrete energy conservation, ensuring an energy-conserving finite-step RTA update. Prescribed lattice heating is introduced by modifying the occupations of carriers already present in the computational domain rather than by continuously creating additional particles. The same framework incorporates surface scattering and explicit material interfaces. Reservoir-driven and locally heated calculations retain the nonlinear Bose--Einstein distribution, whereas the infinite-length in-plane benchmark is treated separately within a periodic-gradient linear-response formulation.

The formulation is evaluated in three stages. First, full-population and deviational calculations with fixed and locally evaluated scattering rates are compared to distinguish the statistical benefit of reference subtraction from the physical effect of temperature-dependent relaxation. Second, cross-plane and in-plane Si films are used to separate reservoir-induced finite-length effects from surface-scattering suppression, while Si/3C-SiC bilayers are used to examine interfacial temperature jumps and the sensitivity to the adopted mixed mismatch model. Finally, localized heating in FinFET-like structures is simulated to determine how substrate composition and interfacial transmission modify the temperature field over a range of heating powers.

\section{Methods}
\label{sec:methods}

\begin{figure*}[htbp]
\centering
\includegraphics[width=0.9\textwidth]{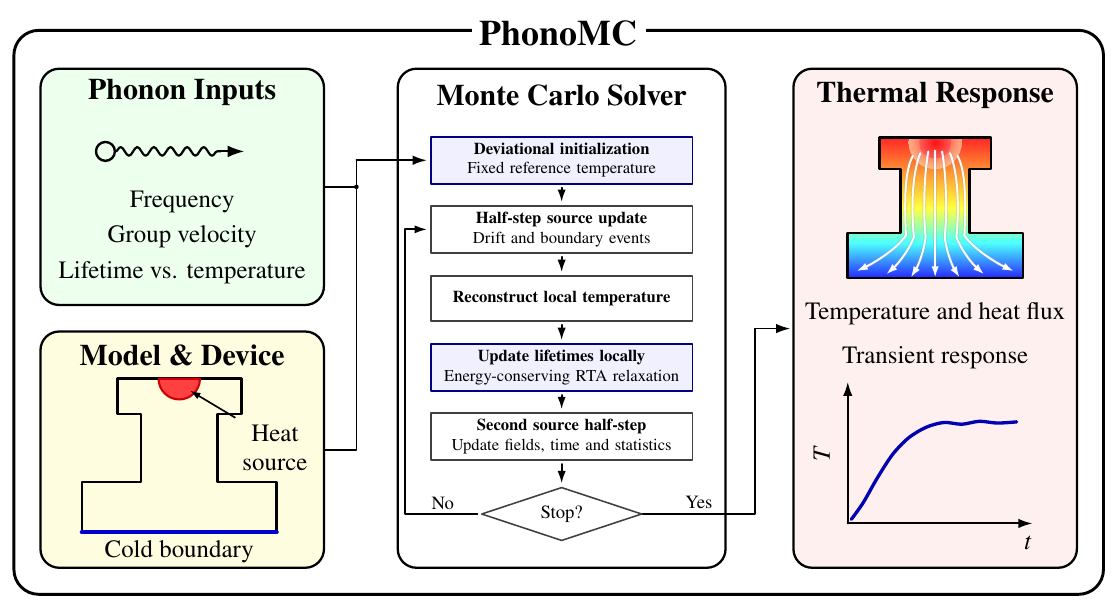}
\caption{PhonoMC framework. Left: phonon properties, device geometry, and thermal driving. Center: carrier initialization, half-step source update, transport and boundary interactions, local temperature reconstruction, energy-conserving relaxation, and the second half-step source update. Right: thermal fields and transient response.}
\label{fig:workflow}
\end{figure*}

PhonoMC solves the mode-resolved phonon Boltzmann transport equation using computational carriers with fixed statistical weights and evolving modal occupations. A fixed equilibrium distribution defines the deviational reference, while the local thermal state is reconstructed from the represented carrier energy. Temperature-dependent scattering, finite-step energy-conserving relaxation, and prescribed lattice heating are treated within the same carrier representation. Figure~\ref{fig:workflow} summarizes the numerical procedure.

\subsection{Transport formulation}
\label{sec:transport_formulation}

Each phonon mode \(\lambda=(\mathbf q,s)\) is characterized by its angular frequency \(\omega_\lambda\), group velocity \(\mathbf v_\lambda\), and temperature-dependent lifetime \(\tau_\lambda(T)\). These quantities are obtained from harmonic and anharmonic lattice-dynamics calculations over the sampled Brillouin zone. Frequencies and group velocities are treated as temperature independent, whereas modal occupations and scattering rates evolve with the thermal state. Within the RTA, the total relaxation rate includes intrinsic phonon scattering together with any separately supplied isotope or defect scattering rates. Normal and Umklapp processes enter through their contributions to the total mode-dependent lifetime.

The mode occupation \(f_\lambda(\mathbf r,t)\) satisfies
\begin{equation}
\frac{\partial f_\lambda}{\partial t}
+\mathbf v_\lambda\cdot\nabla_{\mathbf r}f_\lambda
=
-\frac{f_\lambda-f_\lambda^{\mathrm{eq}}(T_R)}
{\tau_\lambda(T_\tau)}
+S_\lambda^Q,
\label{eq:bte_rta_source}
\end{equation}
where \(S_\lambda^Q\) represents prescribed lattice heating and
\begin{equation}
f_\lambda^{\mathrm{eq}}(T)
=
\left[
\exp\!\left(
\frac{\hbar\omega_\lambda}{k_{\mathrm B}T}
\right)-1
\right]^{-1}
\label{eq:be_distribution}
\end{equation}
is the Bose--Einstein distribution.

Two temperatures enter the collision term. \(T_\tau\) determines the temperature-dependent lifetime \(\tau_\lambda\), whereas \(T_R\) defines the equilibrium distribution toward which the occupations relax. For local-property calculations, \(T_\tau\) is the reconstructed local temperature; for fixed-property calculations it is prescribed. The relaxation temperature \(T_R\) is obtained from the energy-conservation condition
\begin{equation}
\sum_{\lambda\in\mathcal A}
\frac{\hbar\omega_\lambda}
{\tau_\lambda(T_\tau)}
\left[
f_\lambda-f_\lambda^{\mathrm{eq}}(T_R)
\right]
=0,
\label{eq:continuous_collision_energy}
\end{equation}
where \(\mathcal A\) denotes the retained transport-mode set. Since the collision balance is weighted by the mode-dependent rates, \(T_R\) generally differs from the temperature reconstructed from the total energy.

The equilibrium energy density is
\begin{equation}
u^{\mathrm{eq}}(T)
=
\frac{1}{N_{\mathbf q}V_{\mathrm{uc}}}
\sum_{\lambda\in\mathcal A}
\hbar\omega_\lambda
f_\lambda^{\mathrm{eq}}(T),
\label{eq:equilibrium_energy_density}
\end{equation}
where \(N_{\mathbf q}\) is the number of wave vectors in the Brillouin-zone mesh and \(V_{\mathrm{uc}}\) is the primitive-cell volume. The temperature-independent zero-point contribution is omitted. Because \(u^{\mathrm{eq}}(T)\) is monotonic, the local temperature is obtained by numerical inversion of this relation.

Computational carriers sample spatial positions and phonon modes and carry modal occupations \(n_p\). For uniform sampling in a single material, all carriers use the same phase-space weight
\begin{equation}
\mathcal W
=
\frac{N_{\mathrm m}V_\Omega}
{N_p^0N_{\mathbf q}V_{\mathrm{uc}}},
\label{eq:carrier_weight}
\end{equation}
where \(N_{\mathrm m}=|\mathcal A|\) is the number of retained wavevector--branch states, \(V_\Omega\) is the material-domain volume, and \(N_p^0\) is the initial carrier number. A carrier with modal energy \(e_p\) therefore represents the physical energy \(\mathcal W e_p\). In multimaterial calculations, the initial carrier allocation is chosen so that the same \(\mathcal W\) is used in all materials.

Relative to a fixed reference temperature \(T_{\mathrm{ref}}\), carrier \(p\) represents the modal energy deviation
\begin{equation}
e_p^{\mathrm{dev}}
=
\hbar\omega_{\lambda_p}
\left[
n_p-f_{\lambda_p}^{\mathrm{eq}}(T_{\mathrm{ref}})
\right].
\label{eq:particle_deviational_energy}
\end{equation}
For a spatial cell \(g\) with volume \(V_g\),
\begin{align}
    u_g
    &=
    u^{\mathrm{eq}}(T_{\mathrm{ref}})
    +
    \frac{\mathcal W}{V_g}
    \sum_{p\in g} e_p^{\mathrm{dev}},
    \label{eq:grid_energy_density}
    \\
    T_g
    &=
    \left(u^{\mathrm{eq}}\right)^{-1}(u_g),
    \label{eq:local_temperature_update}
    \\
    \mathbf q_g
    &=
    \frac{\mathcal W}{V_g}
    \sum_{p\in g}
    e_p^{\mathrm{dev}}\mathbf v_p.
    \label{eq:average_heat_flux}
\end{align}
For multimaterial calculations, \(u^{\mathrm{eq}}(T)\) is evaluated using the phonon spectrum of the material occupying the cell.

The reference temperature \(T_{\mathrm{ref}}\) defines only the deviational representation. The reconstructed temperature \(T_g\) is obtained from the total represented energy and is used to evaluate temperature-dependent material properties. Finite-temperature calculations retain the full Bose--Einstein distribution. For the full-population reference calculations, carriers sample the total modal occupation directly rather than its deviation from \(T_{\mathrm{ref}}\).

\subsection{Time integration and energy-conserving relaxation}
\label{sec:time_integration}

Each time step consists of a half-source update, carrier transport, local thermal reconstruction, collision, and a second half-source update. During transport, carriers propagate according to their modal group velocities and undergo any boundary or interface interactions encountered along the trajectory. The local temperature is then reconstructed and used to evaluate the scattering rates for the collision step. The spatial mesh is used for thermal reconstruction and cellwise collisions; crossing a cell boundary does not itself constitute a scattering event.

Within cell \(g\), the lifetimes are evaluated at \(T_{\tau,g}\) and held fixed during the collision interval \(\Delta t\). The finite-step relaxation weight of carrier \(p\) is
\begin{equation}
a_p
=
1-\exp\!\left[
-\frac{\Delta t}
{\tau_{\lambda_p}(T_{\tau,g})}
\right].
\label{eq:rta_relaxation_weight}
\end{equation}
For a fixed target occupation, this factor gives the exact single-mode RTA relaxation over \(\Delta t\). The common relaxation temperature \(T_{R,g}\) is determined by requiring zero net energy change during the collision step,
\begin{equation}
\sum_{p\in g}
\hbar\omega_{\lambda_p}a_p
\left[
n_p^*
-
f_{\lambda_p}^{\mathrm{eq}}(T_{R,g})
\right]
=0,
\label{eq:rta_discrete_energy}
\end{equation}
where \(n_p^*\) is the occupation immediately before collision. The occupations are then updated as
\begin{equation}
n_p^{**}
=
(1-a_p)n_p^*
+
a_p f_{\lambda_p}^{\mathrm{eq}}(T_{R,g}).
\label{eq:rta_occupation_update}
\end{equation}
The common phase-space weight cancels from the conservation equation, reducing the determination of \(T_{R,g}\) to a scalar monotonic root solve. Collisions are therefore represented entirely through occupation relaxation, without resampling carrier modes or changing the computational carrier population.

\subsection{Periodic-gradient linear-response formulation}
\label{sec:periodic_gradient}

For the infinite-length in-plane benchmark, a separate linear-response formulation is used with periodic boundaries and an imposed macroscopic temperature gradient \(G=\partial T/\partial x\). The sampled distribution represents the periodic deviation from equilibrium at \(T_{\mathrm{ref}}\), avoiding explicit hot and cold reservoirs. Modal heat capacities and scattering rates are evaluated at the reference temperature.

The modal heat capacity is
\begin{equation}
C_\lambda
=
\hbar\omega_\lambda
\left(
\frac{\partial f_\lambda^{\mathrm{eq}}}{\partial T}
\right)_{T_{\mathrm{ref}}},
\label{eq:modal_heat_capacity}
\end{equation}
and the corresponding energy-driving rate is
\begin{equation}
D_\lambda
=
-C_\lambda v_{\lambda,x}G.
\label{eq:gradient_source}
\end{equation}
The full Brillouin-zone sum of \(D_\lambda\) vanishes by symmetry. Finite carrier sampling does not reproduce this cancellation exactly, so the source increment over a substep \(\delta t\) is projected as
\begin{equation}
\Delta e_p^{\nabla T}
=
\delta t
\left[
D_p
-
C_p
\frac{\sum_{j\in g}D_j}
{\sum_{j\in g}C_j}
\right],
\label{eq:sampled_gradient_projection}
\end{equation}
which removes the finite-sampling net-energy bias while retaining the imposed gradient.

The linearized conserving collision operator is
\begin{align}
    \left.\frac{de_p}{dt}\right|_{\mathrm{coll}}
    &=
    -r_p\left(e_p-C_p\theta_{R,g}\right)
    \label{eq:linear_response_rta}
    \\
    \theta_{R,g}
    &=
    \frac{\sum_{p\in g} r_p e_p}
         {\sum_{p\in g} r_p C_p},
    \qquad
    r_p=\tau_{\lambda_p}^{-1}(T_{\mathrm{ref}})
    \label{eq:linear_response_target}
\end{align}
Here \(e_p=e_p^{\mathrm{dev}}\), and \(\theta_{R,g}\) is the conserving temperature perturbation. The resulting linear collision dynamics are advanced using the exponential action of the operator.

\subsection{Boundary and interface scattering}
\label{sec:boundaries_interfaces}

Carriers propagate along their modal group velocities. Boundary events are processed sequentially within each flight, and the remaining flight time is continued after each interaction. Periodic boundaries map carriers to the paired surface without changing their modal state or occupation. Thermal reservoirs absorb incident carriers and inject new carriers with normal-flux mode weighting and Bose--Einstein occupations at the reservoir temperature. Single-material calculations use absorbed-count replacement, whereas interface calculations use independent incoming-flux injection.

Surface scattering follows the Ziman--Soffer model~\cite{Ziman1960,Soffer1967},
\begin{equation}
p_{\mathrm{spec},\lambda}
=
\exp\!\left[-(2\eta k_\lambda\mu_\lambda)^2\right],
\qquad
\mu_\lambda
=
\frac{|\mathbf v_\lambda\cdot\mathbf n_b|}
{|\mathbf v_\lambda|},
\label{eq:ziman_soffer}
\end{equation}
where \(\eta\) is the RMS roughness, \(k_\lambda\) is the wave-vector magnitude, and \(\mathbf n_b\) is the surface normal. The incidence factor is evaluated from the group-velocity direction. Surface scattering is treated as elastic, so the phonon frequency is conserved during reflection. For specular reflection, the outgoing mode is selected from the frequency-matched states such that its group-velocity direction is closest to the geometrically reflected direction. Diffuse reflection samples frequency-matched outgoing states with normal-flux weighting. This treatment preserves the phonon energy while allowing the propagation direction to change according to the selected boundary-scattering mechanism.

Material interfaces are treated using a mixed mismatch model (MMM)~\cite{zong_mixed_2023}. Acoustic mismatch (AMM)~\cite{little_transport_1959} and diffuse mismatch (DMM)~\cite{swartz_thermal_1989} scattering kernels are combined with weights \(p\) and \(1-p\), respectively. Interface scattering is treated as elastic, so the phonon frequency is conserved upon transmission or reflection. The DMM branch redistributes transmitted or reflected carriers among frequency-matched outgoing states using diffuse flux weighting. The AMM branch uses a scalar acoustic approximation with restricted matching of the wave-vector component parallel to the interface. Polarization conversion is not included in this branch, and optical modes are reflected. The mixing coefficient \(p\) acts on the scattering probabilities and serves as the interface-model parameter.

\subsection{Occupation-based heat-source deposition}
\label{sec:heat_source}

A prescribed volumetric heating rate \(Q(\mathbf r,t)\) introduces the energy increment
\begin{equation}
\Delta E_g^Q
=
\int_{t_a}^{t_b}
\int_{V_g}
Q(\mathbf r,t)\,dV\,dt
\label{eq:heat_source_energy_density}
\end{equation}
into cell \(g\) during a source substep.

PhonoMC deposits this energy by updating the occupations of carriers already present in the cell. For thermal deposition, an auxiliary temperature \(T_g^Q\) is determined from
\begin{equation}
\mathcal W
\sum_{p\in g}
\hbar\omega_{\lambda_p}
\left[
f_{\lambda_p}^{\mathrm{eq}}(T_g^Q)
-
f_{\lambda_p}^{\mathrm{eq}}(T_g)
\right]
=
\Delta E_g^Q .
\label{eq:thermal_source_balance}
\end{equation}
The corresponding occupation increment is
\begin{equation}
\Delta n_p^Q
=
f_{\lambda_p}^{\mathrm{eq}}(T_g^Q)
-
f_{\lambda_p}^{\mathrm{eq}}(T_g).
\label{eq:thermal_source_occupation}
\end{equation}
The computational carrier number is therefore unchanged by heat deposition. For selective excitation, the prescribed energy is instead distributed according to normalized mode weights, so that the spatial and spectral source profiles can be specified independently. The source represents prescribed lattice heating; electronic transport is not solved explicitly.

\subsection{Thermal observables}
\label{sec:observables}

Thermal fields are obtained by averaging cell temperatures and heat fluxes over the selected sampling window. For reservoir-driven transport,
\begin{equation}
\kappa_{\mathrm{eff}}
=
-\frac{\overline{\langle q_\alpha\rangle_\Omega}}
{(T_{\mathrm{right}}-T_{\mathrm{left}})/L_\alpha},
\label{eq:kappa_eff}
\end{equation}
where \(\langle\cdot\rangle_\Omega\) denotes the volume average, the overbar denotes temporal averaging, and \(L_\alpha\) is the reservoir separation along the transport direction. Because the full reservoir temperature difference is used, \(\kappa_{\mathrm{eff}}\) includes the reservoir-associated temperature drops.

An interior-gradient estimate is defined as
\begin{equation}
\kappa_{\mathrm{fit}}
=
-\frac{\overline{\langle q_\alpha\rangle_\Omega}}
{G_{\mathrm{fit}}},
\label{eq:kappa_fit}
\end{equation}
where \(G_{\mathrm{fit}}\) is obtained from a linear fit to the interior temperature profile. This estimate excludes the reservoir temperature jumps and characterizes the interior transport response.

For periodic-gradient calculations,
\begin{equation}
\kappa_\parallel(H,\eta;L\rightarrow\infty)
=
-\frac{\overline{\langle q_x\rangle_\Omega}}{G},
\label{eq:kappa_infinite}
\end{equation}
which gives the infinite-length in-plane conductivity at film thickness \(H\) and surface roughness \(\eta\).

The interfacial thermal resistance is
\begin{equation}
R_{\mathrm{int}}
=
\frac{\Delta T_{\mathrm{int}}}{\overline{q_n}},
\label{eq:interface_resistance}
\end{equation}
where \(\Delta T_{\mathrm{int}}\) is obtained by extrapolating the averaged temperature profiles on either side of the interface and \(\overline{q_n}\) is the corresponding normal heat flux. Locally heated structures are characterized by their temperature-rise and heat-flux fields.

Energy conservation is monitored using the domain-energy change, prescribed heat input, and net reservoir exchange. Statistical variability is evaluated from independent realizations or temporal blocks, as specified for each result.

\section{Results and discussion}

\subsection{Sampling convergence and temperature-dependent scattering}

We use cross-plane transport through a 100-nm Si film to separate the effects of variance reduction from those of temperature-dependent scattering. The cold reservoir is fixed at 300~K, while the hot-reservoir temperature is varied to change the imposed temperature difference. All three calculations use the same phonon properties, geometry, and RTA model.

\begin{figure}[!htbp]
    \centering
    \includegraphics[width=0.95\linewidth]{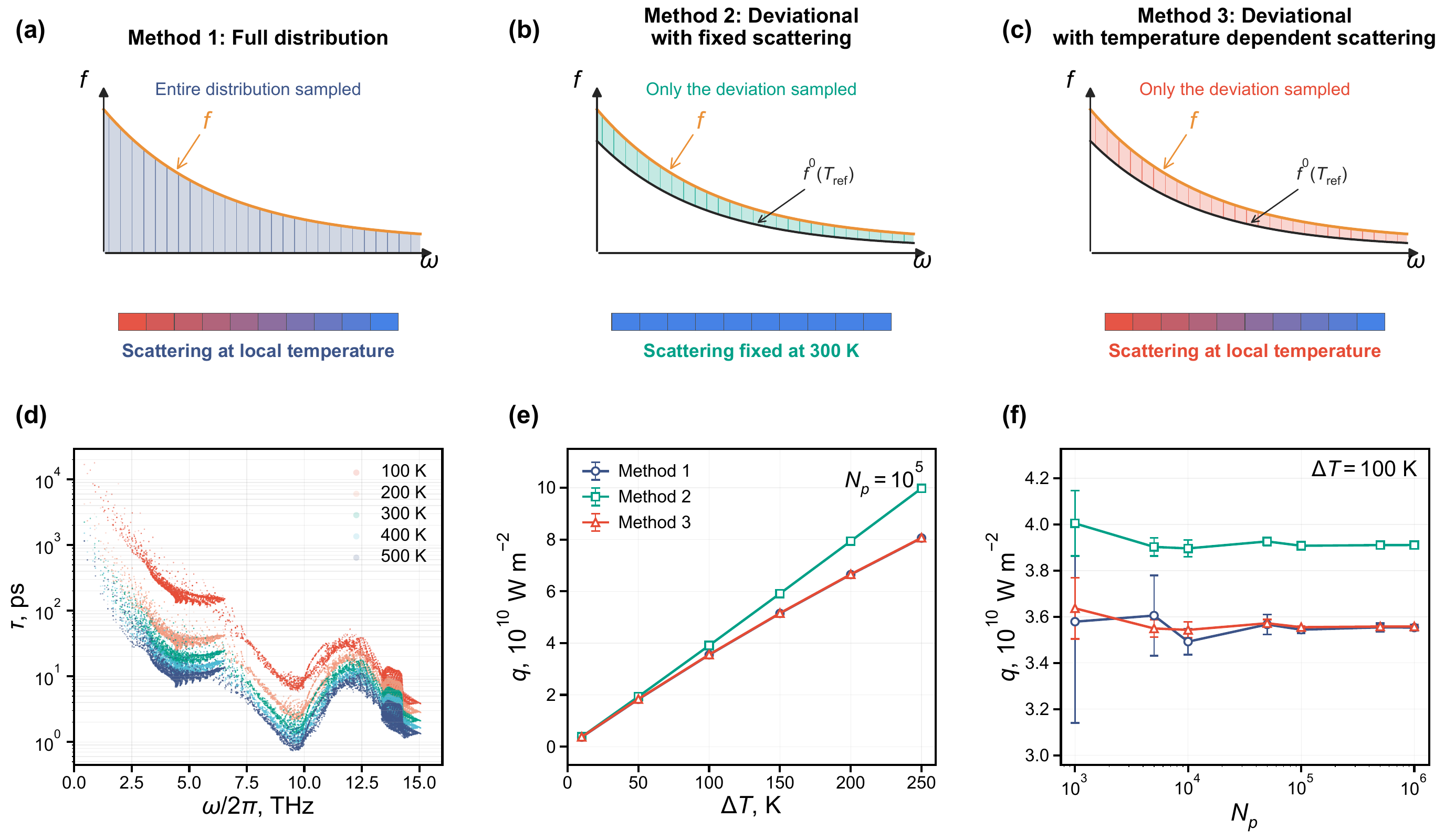}
    \caption{
    Comparison of sampling convergence and temperature-dependent scattering
    in RTA calculations of cross-plane heat transport through a 100-nm Si film.
    (a)--(c) Schematic representations of Method 1 (full distribution),
    Method 2 (deviational with fixed scattering), and Method 3
    (deviational with temperature-dependent scattering), respectively.
    Shading indicates the sampled distribution or deviation.
    Both deviational methods use a fixed reference temperature of
    $T_{\mathrm{ref}}=300~\mathrm{K}$.
    Relaxation times are evaluated at the local temperature in Methods 1
    and 3, and fixed at 300~K in Method 2, as illustrated by the colored strips.
    (d) Mode-resolved phonon relaxation times versus frequency at selected temperatures.
    (e) Time-averaged heat flux versus reservoir temperature difference
    at $N_p=10^5$.
    (f) Carrier-number convergence of the time-averaged heat flux
    at $\Delta T=100~\mathrm{K}$.
    The cold reservoir is maintained at 300~K, with periodic boundaries
    in the transverse directions.
    Symbols denote means over five independent realizations,
    each averaged over 0.75--1~ns; error bars indicate one sample
    standard deviation across these realizations.
    Lines connect the symbols, and the legend in (e) also applies to (f).
    }
    \label{fig:MC_convergence}
\end{figure}

Figures~\ref{fig:MC_convergence}(a)--(c) define the three formulations. Method 1 samples the full phonon distribution and evaluates the relaxation times at the reconstructed local temperature. Method 2 uses a deviational representation with a fixed reference temperature \(T_{\mathrm{ref}}=300~\mathrm{K}\) and also fixes the relaxation times at 300~K. Method 3 uses the same fixed reference but evaluates the relaxation times from the local temperature. The comparison between Methods 1 and 3 therefore isolates the effect of reference subtraction, whereas the comparison between Methods 2 and 3 isolates the effect of fixing the scattering rates. In both deviational formulations, the reference distribution remains unchanged as the local temperature evolves.

The mode-resolved relaxation times in Fig.~\ref{fig:MC_convergence}(d) decrease with increasing temperature over most of the spectrum, so hotter regions experience stronger phonon scattering. This temperature dependence becomes increasingly important as the imposed reservoir difference grows. Methods 1 and 3 give nearly identical mean heat fluxes over the investigated range, whereas Method 2 produces a systematically larger value. Relative to Method 3, the excess heat flux is approximately 1.0\% at \(\Delta T=10~\mathrm{K}\), 9.9\% at 100~K, and 23.7\% at 250~K [Fig.~\ref{fig:MC_convergence}(e)]. Fixing the scattering rates at the cold-reservoir temperature therefore increasingly overestimates transport as the temperature field becomes more nonuniform. Importantly, this bias is not introduced by the deviational representation itself: the equilibrium reference can remain fixed while the scattering rates are updated from the local temperature.

Figure~\ref{fig:MC_convergence}(f) compares the carrier-number convergence at \(\Delta T=100~\mathrm{K}\). Methods 1 and 3 approach the same mean heat flux as the carrier count increases, but Method 3 shows substantially smaller run-to-run variation. At \(N_p=10^5\), the sample standard deviation of Method 3 is approximately 3.3 times smaller than that of Method 1, while at the largest carrier count their mean heat fluxes differ by only about 0.13\%. Method 2 also exhibits small statistical fluctuations, but converges to a different heat flux because the scattering rates remain fixed. Increasing the carrier count therefore reduces stochastic sampling error but does not remove the bias associated with the fixed-scattering approximation.

These comparisons separate two effects that are often coupled in deviational Monte Carlo calculations. Reference subtraction primarily improves statistical sampling, whereas local lifetime evaluation changes the transport solution when the temperature excursion is large. Method 3 retains the variance reduction of the deviational representation while preserving the temperature dependence of the scattering rates.

\subsection{Size-dependent and interfacial thermal transport}

\begin{figure}[!htbp]
\centering
\includegraphics[width=0.95\linewidth]{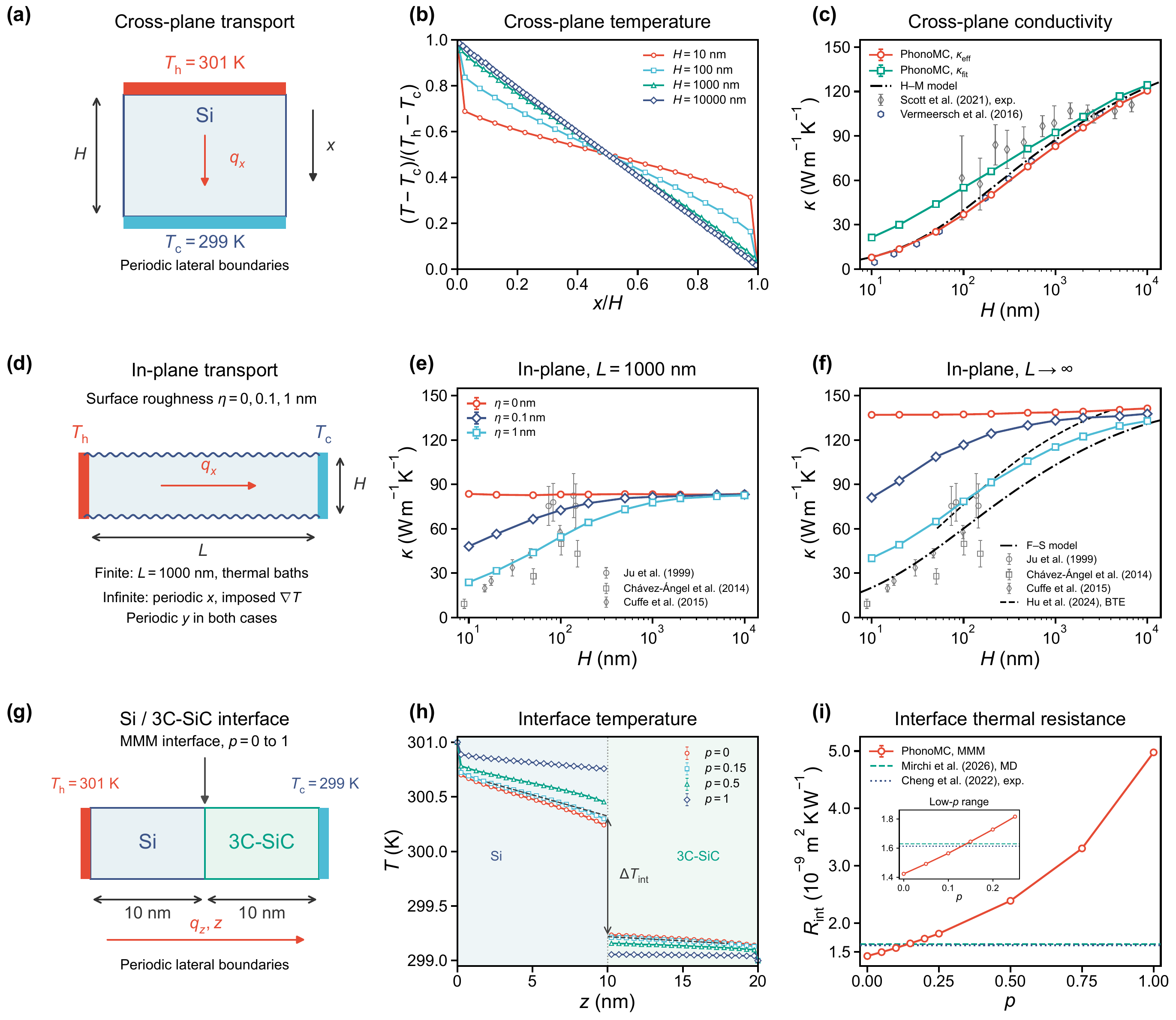}
\caption{Silicon thin-film and Si/3C-SiC interfacial transport near 300~K using RTA. (a,b) Cross-plane geometry and normalized temperature profiles with reservoirs at 301 and 299~K. Symbols denote cell-center temperatures; line endpoints represent the prescribed reservoir temperatures. (c) Cross-plane \(\kappa_{\mathrm{eff}}\) and \(\kappa_{\mathrm{fit}}\), compared with experiments~\cite{scott_simultaneous_2021}, numerical results~\cite{vermeersch_cross-plane_2016}, and the H--M model~\cite{hua_semi-analytical_2015}. (d) In-plane geometry. (e,f) In-plane conductivity for \(L=1000~\mathrm{nm}\) and \(L\rightarrow\infty\), respectively, at surface roughness \(\eta=0\), 0.1, and 1~nm, compared with experiments~\cite{ju_phonon_1999,chavez-angel_reduction_2014,cuffe_reconstructing_2015}. Panel (f) also includes numerical results~\cite{hu_ultra-efficient_2024} and the diffuse F--S model~\cite{fuchs_conductivity_1938,sondheimer_mean_2001}. (g) A 10~nm Si/10~nm 3C-SiC bilayer with an MMM interface; \(p\) denotes the AMM mixing fraction. (h) Temperature profiles for selected \(p\), with interface extrapolations shown for \(p=0.15\). (i) Interfacial thermal resistance versus \(p\), compared with quantum-corrected MD~\cite{mirchi_interfacial_2026} and experimental data~\cite{cheng_high_2022}; the inset enlarges the low-\(p\) region. Simulation error bars show standard deviations across five temporal blocks in (c,e,f) and three independent realizations in (h,i). Literature data are shown under their reported sample conditions.}
\label{fig:si_film_validation}
\end{figure}

Figure~\ref{fig:si_film_validation} summarizes the thin-film and interface calculations near 300~K. Material-specific mode-resolved properties are used for Si and 3C-SiC, with the same RTA treatment throughout, allowing the effects of finite transport length, surface scattering, and interfacial transmission to be examined within a consistent transport model.

For cross-plane transport, the reservoir separation is equal to the film thickness \(H\), with periodic boundaries in the lateral directions [Fig.~\ref{fig:si_film_validation}(a)]. At small \(H\), the temperature profiles show pronounced jumps between the reservoirs and the film [Fig.~\ref{fig:si_film_validation}(b)]. These jumps account for a substantial fraction of the imposed temperature difference and weaken as the film becomes thicker.

The same behavior appears in the effective conductivity [Fig.~\ref{fig:si_film_validation}(c)]. The reservoir-based value, \(\kappa_{\mathrm{eff}}\), increases from approximately 7.8 to \(120.5~\mathrm{W\,m^{-1}K^{-1}}\) as \(H\) increases from 10 to 10,000~nm, approaching the bulk RTA value of \(141.7~\mathrm{W\,m^{-1}K^{-1}}\). The interior-gradient estimate, \(\kappa_{\mathrm{fit}}\), is larger because the fitted gradient excludes the reservoir temperature jumps. The difference between \(\kappa_{\mathrm{eff}}\) and \(\kappa_{\mathrm{fit}}\) decreases with increasing thickness as the reservoir contribution becomes less important. The calculated thickness dependence is consistent with the trends reported experimentally~\cite{scott_simultaneous_2021} and numerically~\cite{vermeersch_cross-plane_2016}, and is also compared with the Hua--Minnich spectral suppression model~\cite{hua_semi-analytical_2015} evaluated using the same phonon properties.

The in-plane calculations separate finite transport-length effects from surface scattering [Fig.~\ref{fig:si_film_validation}(d)--(f)]. Surface specularity is described by the Ziman--Soffer model~\cite{Ziman1960,Soffer1967} with RMS roughness \(\eta=0\), 0.1, and 1~nm. For \(L=1000~\mathrm{nm}\), the conductivity remains below the bulk value even for perfectly specular surfaces, reaching approximately \(84~\mathrm{W\,m^{-1}K^{-1}}\). This reduction is caused by the finite reservoir separation rather than by surface scattering. Increasing \(\eta\) introduces an additional suppression that is strongest in the thinnest films. At larger thicknesses, the rough-surface results approach the specular finite-length curve because the reservoir-induced length effect remains.

The periodic-gradient calculation removes the finite reservoir separation and isolates the surface contribution. In this limit, the specular result approaches the bulk conductivity, whereas rough surfaces retain a pronounced thickness dependence. The diffuse Fuchs--Sondheimer model~\cite{fuchs_conductivity_1938,sondheimer_mean_2001}, evaluated using the same phonon spectrum, gives the corresponding fully diffuse reference. The calculated trends are compared with the experimental data of Refs.~\cite{ju_phonon_1999,chavez-angel_reduction_2014,cuffe_reconstructing_2015} and the numerical results of Ref.~\cite{hu_ultra-efficient_2024}.

We next consider cross-interface transport in a 10~nm Si/10~nm 3C-SiC bilayer between reservoirs at 301 and 299~K [Fig.~\ref{fig:si_film_validation}(g)]. Interfacial scattering is described by the MMM, with \(p\) controlling the AMM contribution to the scattering kernel. A clear temperature discontinuity develops at the interface [Fig.~\ref{fig:si_film_validation}(h)]. Extrapolating the temperature profiles on the two sides gives the interfacial jump \(\Delta T_{\mathrm{int}}\), from which \(R_{\mathrm{int}}=\Delta T_{\mathrm{int}}/q_z\) is evaluated.

For the present MMM implementation, \(R_{\mathrm{int}}\) increases from approximately \(1.42\) to \(4.98\times10^{-9}~\mathrm{m^2\,K\,W^{-1}}\) as \(p\) varies from 0 to 1 [Fig.~\ref{fig:si_film_validation}(i)]. The quantum-corrected MD result of Mirchi et al.~\cite{mirchi_interfacial_2026} and the experimental value reported by Cheng et al.~\cite{cheng_high_2022} are approximately \(1.63\) and \(1.61\times10^{-9}~\mathrm{m^2\,K\,W^{-1}}\), respectively. Interpolation of the simulated curve gives \(p\approx0.14\) for the MD result and \(p\approx0.13\) for the experimental value. Because \(p\) is calibrated here against available Si/3C-SiC data rather than independently predicted, we use \(p=0.14\) as a representative interface parameter in the subsequent device calculations.

\subsection{Localized heating in FinFET structures}

\begin{figure}[htbp]
\centering
\includegraphics[width=0.95\linewidth]{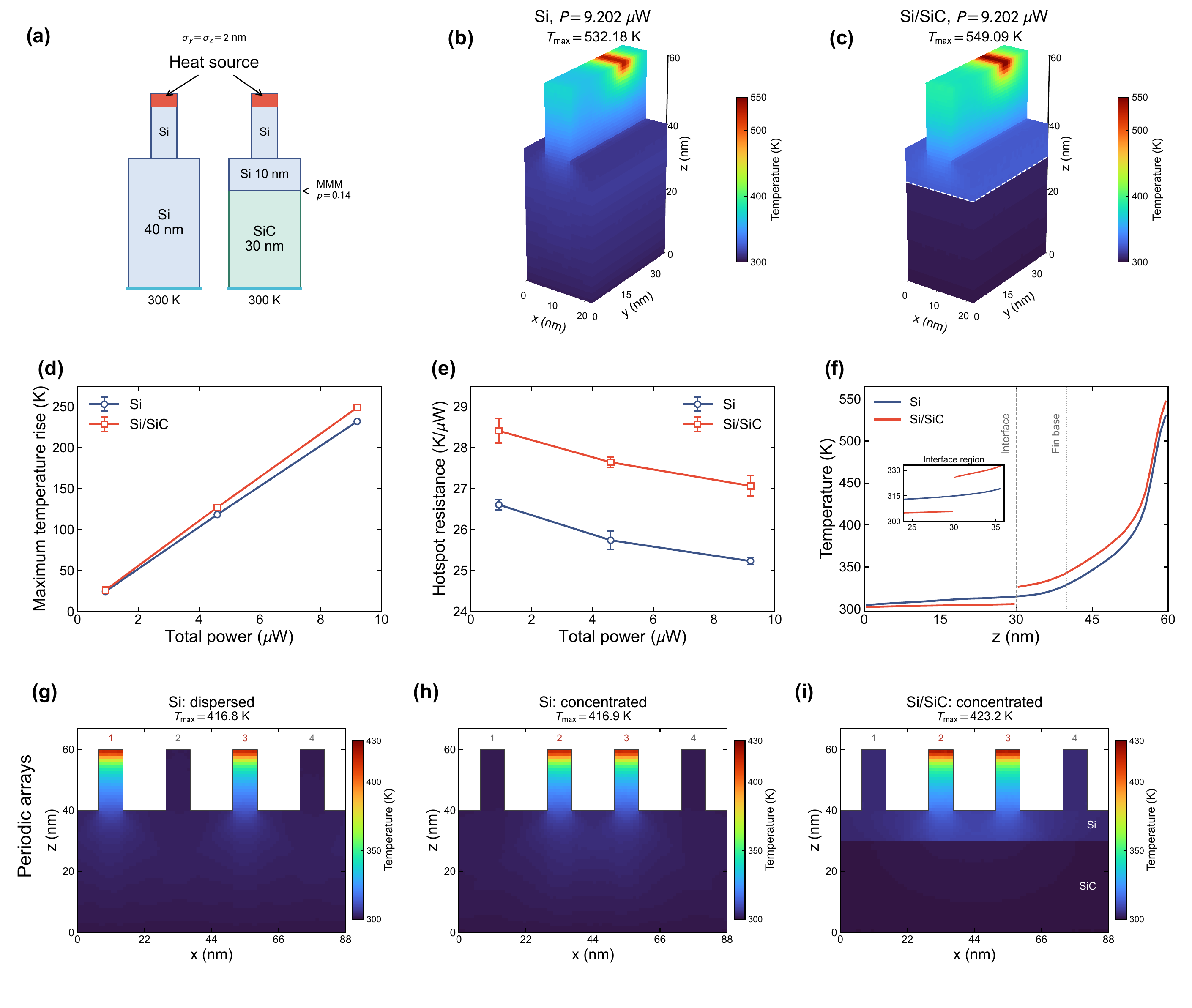}
\caption{\textbf{Localized heating in Si and Si/SiC FinFET structures and periodic arrays.} (a) Device geometries drawn to scale, with a 40~nm Si substrate or 10~nm Si above 30~nm SiC. The bottom reservoir is maintained at 300~K, and the interface uses MMM with \(p=0.14\). (b,c) Single-device temperature fields at \(P=9.202~\mu\mathrm{W}\). (d) Maximum temperature rise and (e) hotspot thermal resistance versus total power. Error bars indicate the standard deviation of five temporal-block temperatures at the fixed hotspot location. (f) Vertical temperature profiles through the heat-source center, with an interface-region inset. (g--i) Temperature maps in the \(y=25\)~nm cross section for dispersed heating in Si, concentrated heating in Si, and concentrated heating in Si/SiC, respectively. Dispersed and concentrated heating activate fins 1 and 3, and fins 2 and 3, respectively, at the same total power of \(9.202~\mu\mathrm{W}\). The array maps use a common 200~ps averaging window and identical color scales. White gaps denote regions outside the material, and the dashed line marks the Si/SiC interface. Annotated maxima are obtained from the full three-dimensional time-averaged temperature fields.}
\label{fig:fet_summary}
\end{figure}

We next apply PhonoMC to localized heating in three-dimensional FinFET structures and examine the effects of substrate composition, interface transmission, and heat-source arrangement. Figure~\ref{fig:fet_summary}(a) compares two structures with the same external dimensions: a Si fin on a 40~nm Si substrate and a Si fin on a layered substrate consisting of 10~nm Si above 30~nm 3C-SiC. The fin is 8~nm wide and 20~nm high, and the single-device periodic cell spans \(22\times36\)~nm\(^2\) in the lateral directions. Periodic boundaries are imposed along \(x\) and \(y\), and the bottom reservoir is held at 300~K. The Si/SiC interface is described by the MMM with \(p=0.14\). The relaxation times are evaluated from the reconstructed local temperature.

The prescribed heat source is placed near the top of the fin. Heating is uniform across the fin width and Gaussian in the \(y\)--\(z\) plane, with \(\sigma_y=\sigma_z=2\)~nm. Peak volumetric heat-generation rates of \(10^{19}\), \(5\times10^{19}\), and \(10^{20}~\mathrm{W\,m^{-3}}\) correspond to total powers of approximately 0.920, 4.601, and 9.202~\(\mu\)W, respectively. The calculations therefore probe phonon transport under controlled localized lattice heating.

Figures~\ref{fig:fet_summary}(b,c) show the single-device temperature fields at the highest power. In both structures, the maximum temperature is located near the heat source and decreases toward the bottom reservoir. The peak temperature is 532.2~K for the Si substrate and 549.1~K for the Si/SiC substrate. Replacing the lower part of the Si substrate with 3C-SiC therefore raises the hotspot temperature by about 16.9~K under the present interface conditions. Despite the higher bulk thermal conductivity of 3C-SiC, the layered structure is hotter because the heat-removal path also includes transport through the confined Si region and transmission across the Si/SiC interface.

The power dependence is shown in Figs.~\ref{fig:fet_summary}(d,e). We define the hotspot thermal resistance as \(R_{\mathrm{hot}}=(T_{\max}-T_0)/P\), with \(T_0=300\)~K. Over the investigated power range, \(R_{\mathrm{hot}}\) changes only modestly, from approximately 26.6 to 25.2~K/\(\mu\)W for Si and from 28.4 to 27.1~K/\(\mu\)W for Si/SiC. The layered structure remains more resistive at all three powers, with a temperature rise about 7\% larger than that of the corresponding Si structure. Here \(R_{\mathrm{hot}}\) characterizes the thermal response of the complete device geometry, including the source distribution, interfaces, and temperature-dependent phonon scattering.

The vertical profiles in Fig.~\ref{fig:fet_summary}(f) clarify the origin of the higher hotspot temperature in the layered structure. A clear temperature discontinuity appears at the Si/SiC interface at \(z=30\)~nm, and the Si region above the interface remains hotter than in the all-Si device. At the highest power, extrapolation of the two adjacent profiles gives an interface temperature jump of approximately 20~K. This additional temperature drop reduces heat removal from the upper Si region and is consistent with the larger \(T_{\max}\) of the Si/SiC structure.

We then extend the geometry to four fins arranged along \(x\) with a center-to-center spacing of 22~nm. The periodic supercell spans \(88\times36\)~nm\(^2\), with the same substrate thickness and bottom-reservoir condition as the single-fin calculations. In the dispersed configuration, fins 1 and 3 are heated; in the concentrated configuration, the adjacent fins 2 and 3 are heated. Each active fin uses a peak heat-generation rate of \(5\times10^{19}~\mathrm{W\,m^{-3}}\), so both configurations inject the same total power of approximately 9.202~\(\mu\)W.

Figures~\ref{fig:fet_summary}(g--i) compare the resulting temperature fields over the same 200~ps averaging window. In the all-Si array, the dispersed and concentrated configurations give nearly identical peak temperatures, 416.8 and 416.9~K, respectively. The difference is much smaller than the temporal fluctuations, so no systematic increase in \(T_{\max}\) is resolved when the two active fins are placed next to each other. The spatial temperature fields are nevertheless different, with the concentrated case producing a more localized warm region around the adjacent sources. For the concentrated configuration, replacing the lower substrate with SiC raises the peak temperature to approximately 423.2~K, about 6.4~K above the corresponding all-Si case, and leaves the Si region above the interface systematically warmer.

These device calculations show how the occupation-based source treatment can be combined with local temperature-dependent scattering and explicit interfaces in a three-dimensional geometry. The results are intended as phonon-transport calculations under prescribed lattice heating; quantitative electrothermal prediction of a specific transistor would additionally require realistic contacts, dielectric regions, and an electronically determined heat-generation profile.

\section{Conclusions}

We developed PhonoMC, an occupation-based deviational Monte Carlo method for mode-resolved phonon transport under finite temperature variations. The central feature of the formulation is that the deviational reference, the local thermal state, and the conserving relaxation state are treated separately. A fixed equilibrium reference is retained for variance reduction, while the local temperature reconstructed from the represented energy is used to evaluate temperature-dependent scattering rates. Collisions are advanced with a finite-step RTA update in which the relaxation temperature is determined from discrete energy conservation. Prescribed lattice heating is introduced through changes in carrier occupations, so sustained energy injection does not require growth of the computational carrier population.

The numerical results separate the roles of statistical sampling and temperature-dependent scattering. For cross-plane transport through a 100-nm Si film, the deviational formulation reproduces the full-population heat flux with substantially smaller run-to-run fluctuations. Fixing the scattering rates at 300~K, however, produces a systematic heat-flux overestimate that reaches 23.7\% at a reservoir temperature difference of 250~K. Thin-film calculations further distinguish reservoir-induced finite-length effects from surface-scattering suppression, while Si/3C-SiC bilayers show how the interfacial thermal resistance varies with the adopted transmission kernel.

Application to localized heating in FinFET structures demonstrates the effect of combining local temperature-dependent scattering with explicit material interfaces. For the selected geometry and interface parameter, replacing the lower Si substrate with 3C-SiC raises the peak temperature from 532.2 to 549.1~K at the highest single-device power, despite the higher bulk thermal conductivity of 3C-SiC. The increase originates from the full heat-removal path, including transport through the confined Si region and transmission across the Si/SiC interface. In the four-fin array, changing the spatial arrangement of the heat sources modifies the temperature field without producing a resolved change in the peak temperature for the all-Si structure over the sampled window.

The present implementation establishes a basis for extending PhonoMC beyond mode-dependent RTA transport. Natural next steps include momentum-conserving collision models, more detailed interface-scattering treatments, coupling to electron transport and electron--phonon energy exchange, and simulations of realistic multi-material device stacks. These developments would allow the same occupation-based formulation to address increasingly complex nonequilibrium thermal transport problems without giving up mode resolution or local temperature dependence.

\section*{CRediT authorship contribution statement}

Shixian Liu: Conceptualization, Methodology, Software, Investigation, Formal analysis, Visualization, Writing -- original draft, Writing -- review \& editing.
Fei Yin: Software, Investigation, Formal analysis.
Gang Wang: Methodology, Investigation, Formal analysis.
Bin Liu: Methodology, Formal analysis.
Ge Zhang: Investigation, Validation.
Alexander A. Barinov: Supervision, Writing -- review \& editing.
Ke Xu: Methodology, Formal analysis, Writing -- review \& editing.

\section*{Declaration of competing interest}

The authors have no conflicts to disclose.

\section*{Data availability}

The input files, processed numerical data, and final figure files used to generate the main figures in this study are available on Zenodo at
\url{https://doi.org/10.5281/zenodo.22817255}.
Additional simulation parameters are provided in the Supplementary Information.

\section*{Code availability}
The PhonoMC source code used in this study is publicly available at:
\url{https://github.com/lyushisyan/PhonoMC/releases/tag/v1.0}.

\section*{Acknowledgements}

S.L. and F.Y. acknowledge financial support from the China Scholarship
Council under Grant Nos. 202308090243 and 202408090635, respectively.
The authors thank Xin Wu, Roman Anufriev, and Vladimir I. Khvesyuk
for helpful discussions.

\bibliographystyle{elsarticle-num} 
\bibliography{my-refs}

\end{document}